\documentclass[journal]{IEEEtran}
\usepackage[columnwise]{lineno}
\usepackage{cite}
\ifCLASSINFOpdf
  \usepackage[pdftex]{graphicx}
\else
  \usepackage[dvips]{graphicx}
\fi
\usepackage{amsmath}
\usepackage{amssymb}
\usepackage{url}
\usepackage{hyperref}
\hypersetup{
 colorlinks=True,
}
\usepackage{subfigure}
\usepackage{booktabs}
\usepackage{multirow}
\renewcommand{\vec}[1]{\mathbf{#1}}
\newcommand{\cmmnt}[1]{}

\usepackage[table,xcdraw, dvipsnames]{xcolor}
\usepackage{float}

\begin{document}
%
\title{Towards Domain Invariant Heart Sound Abnormality Detection using Learnable Filterbanks}

\author{Ahmed~Imtiaz~Humayun,$^{1}$ ~\IEEEmembership{Student Member,~IEEE,}
        Shabnam~Ghaffarzadegan,$^{2}$ 
        Md.~Istiaq~Ansari,$^{1}$ 
        Zhe~Feng,$^{2}$~and 
        Taufiq~Hasan,$^{1}{^{\dagger}}$~\IEEEmembership{Member,~IEEE}
\thanks{$^{1}$Taufiq Hasan, Ahmed Imtiaz Humayun and Md.~Istiaq~Ansari are affiliated with mHealth Laboratory, Department of Biomedical Engineering, Bangladesh University of Engineering and Technology (BUET), Dhaka - 1205, Bangladesh. Email: {\tt\scriptsize taufiq@bme.buet.ac.bd}. 
}
\thanks{$^{2}$Shabnam Ghaffarzadegan and Zhe Feng are with the Human Machine Interaction Group-2, Robert Bosch Research and Technology Center (RTC), Sunnyvale, CA - 94085, USA. Email: {\tt\scriptsize \{shabnam.ghaffarzadegan,zhe.feng2\}@us.bosch.com}}
\thanks{$^{*}$ This work was supported by Robert Bosch Research and Technology Center (Sunnyvale, CA - 94085, USA) and Brain Station 23 (Dhaka - 1212, Bangladesh). The TITAN Xp GPU used for this work was donated by the NVIDIA Corporation.}
}

\markboth{IEEE Journal of Biomedical and Health Informatics,~Vol.~xx, No.~xx, xxxx~2020}%
{Shell \MakeLowercase{\textit{et al.}}: Bare Demo of IEEEtran.cls for IEEE Journals}



\maketitle
\begin{abstract}
\emph{Objective:} Cardiac auscultation is the most practiced non-invasive and cost-effective procedure for the early diagnosis of heart diseases. While machine learning based systems can aid in automatically screening patients, the robustness of these systems is affected by numerous factors including the stethoscope/sensor, environment, and data collection protocol.
This paper studies the adverse effect of domain variability on heart sound {abnormality detection} and develops strategies to address this problem. \emph{Methods:} We propose a novel Convolutional Neural Network (\emph{CNN}) layer, consisting of time-convolutional (\emph{tConv}) units, that emulate Finite Impulse Response (\emph{FIR}) filters. The filter coefficients can be updated via backpropagation and be stacked in the front-end of the network as a learnable filterbank. 
\emph{Results:} 
{On publicly available multi-domain datasets}, the proposed method surpasses the top-scoring systems found in the literature for heart sound {abnormality detection (a binary classification task). We utilized sensitivity, specificity, F-1 score and Macc (average of sensitivity and specificity) as performance metrics}.
Our systems achieved relative improvements of up to 11.84\% in terms of MAcc, compared to state-of-the-art methods.
\emph{Conclusion:} 
The results demonstrate the effectiveness of the proposed learnable filterbank CNN architecture in achieving robustness towards sensor/domain variability in PCG signals.
\emph{Significance:} The proposed methods pave the way for deploying automated cardiac screening systems in diversified and underserved communities. 
\end{abstract}

\begin{IEEEkeywords}
Heart sound classification, learnable filterbank, domain adaptation.
\end{IEEEkeywords}

%
\IEEEpeerreviewmaketitle

\section{Introduction}


\IEEEPARstart{C}{ardiovascular} diseases (CVDs) cause an estimated 17.9 million deaths worldwide while more than 75\% of these deaths occur in low- and medium-income countries \cite{whofact}. The severe shortage of trained physicians and health workers in these underserved communities who can perform cardiac auscultation \cite{alam2010cardiac} exacerbates the condition as early diagnosis of CVDs become unlikely. This demands the development of machine learning-based assistive technologies for cardiac screening. With the advent of smartphones with increased computational capabilities, mobile cardiac auscultation frameworks using wireless digital stethoscopes can thus have a significant impact on public health if they are deployed in the point-of-care locations.


\begin{figure}[t]
\includegraphics[width=\linewidth]{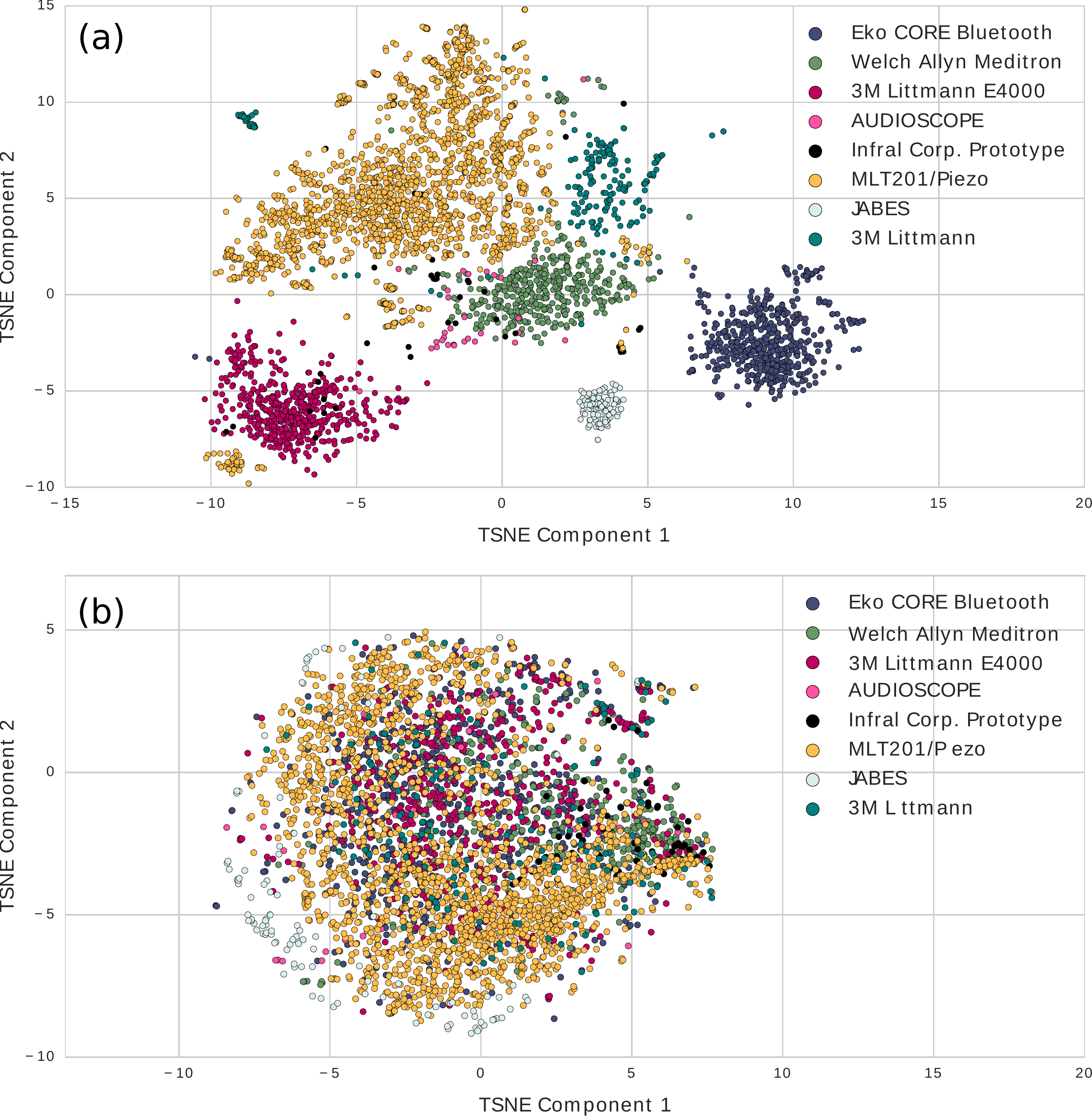}
\centering
\vspace{-2mm}
\caption{t-Stochastic Neighbour Embedding (t-SNE) visualization of 2016 Physionet/CinC heart sound data subsets acquired using various stethoscopes. (a) $6373$ standard computational paralinguistics (ComParE) features comprising of functionals calculated over low level descriptor contours \cite{compare1st}. (b) Flatten layer features of our proposed method with a learnable auditory filterbank front-end. 
Visible clusters corresponding to each of the domains are present in (a) while our proposed multi-source adaptation method normalizes such differences in (b). Details are available in sections \ref{chall} \& \ref{results}.}
\label{tsne}
\vspace{-3mm}
\end{figure}

While geographically distributed cardiac screening systems can change the rural healthcare landscape, it brings newer challenges regarding the acquisition and diagnostic performance. Currently available commercial {and proprietary} systems are reliant on the homogeneity of acquired data. Scaling up such solutions require them to be {more generic} and robust towards heterogeneous sources, i.e. data acquired from different stethoscopes (especially of lower cost) and diverse environments. {In our previous work, we observed that} the frequency characteristics of the stethoscope or sensor used for recording can cause machine learning models to be biased towards majority sources of training data \cite{humayunFIR,Humayun2018}. The visible clusters in Fig \ref{tsne}-(a) corresponding to different stethoscope models prove that feature distributions can be significantly different depending on which domain the data is from. This is a long-standing challenge for any low resource, multi-source, deep learning task; therefore, it needs to be addressed for PCG as well.


In the past few decades, the study of \emph{HS} abnormality detection has been mostly focused on PCG segmentation \cite{messner2018heart},
and  binary classification of PCG as pathologic (\emph{Abnormal}) or physiologic (\emph{Normal}) \cite{liu2016open}. 
In 2016, the Physionet/CinC Challenge was organized and an archive of $4430$ PCG recordings, acquired using seven different stethoscope models, was released for binary classification of heart sounds (\emph{HS}).
Notable features reported for this dataset includes, time, frequency and statistical features \cite{homsi2017}, Mel-frequency Cepstral Coefficients (MFCC) \cite{bobillo2016}, and Continuous Wavelet Transform (CWT) \cite{kay2017}.
Classifiers like SVM \cite{whitaker2017}, k-Nearest Neighbor (k-NN) \cite{bobillo2016}, Multilayer Perceptron (MLP) \cite{kay2017,zabihi2016} and Random Forest \cite{homsi2017}, deep learning approaches with 1D \& 2D CNNs \cite{potes2016ensemble,maknickas2017}, and Recurrent Neural Network (RNN) \cite{yang2016classification} based architectures were employed in the challenge submissions. The winning algorithm, similar to a good number of other submissions, proposed an ensemble; a static filter front-end 1D-CNN model combined with an Adaboost-Abstain classifier using a threshold-based voting algorithm.
In 2018, the INTERSPEECH ComParE Heart Beats Sub-Challenge {\cite{schuller2018interspeech}} introduced a new dataset for multi-class \emph{HS} classification according to severity collected with a single stethoscope model.
The best algorithm among the submissions was made by Gabor et al. \cite{Gabor2018} which outperformed deep learning methods by utilizing segment level static features. Presented methods in both of these competitions focused solely on the classification metric of interest. 
However, to the best of our knowledge, the issue of domain mismatch between stethoscope models has not been addressed in the literature for any dataset to date.

In our previous work \cite{Humayun2018}, we explored an adaptation technique to learn transferable representations across corpora {for heart sound abnormality detection}. We experimentally observed that a sequence-to-sequence autoencoder trained with data from different domains (the 2016 Physionet dataset and the 2018 ComParE dataset) would learn to discriminate between the corpus instead of the pathological classes, i.e., the distribution of features was significantly different depending on data source.
In this work, we extrapolate on our findings of stethoscope dependent variances and quantify domain mismatch for PCG signals. {Following our previous work, we exclusively focus on heart sound classification between two classes, namely normal and abnormal.}
We expand the scope of our learnable filterbank approach originally proposed in \cite{humayunFIR} incorporating novel variants of \emph{tConv} layers including Type I-IV FIR and Gammatone auditory filterbanks. In addition, we propose a training regime within each mini-batch termed Domain Balanced Training (DBT), to specifically address domain differences. 
The DBT method is agnostic of the type of data and therefore can be applied in other applications as a supervised domain adaptation method. 

The remainder of this paper is organized as follows. In Section \ref{pcg} \& \ref{dataset}, we discuss the spectro-temporal characteristics of \emph{HS} and the available datasets. In Sec. \ref{chall} \& \ref{method}, we elaborate on the challenges with the PCG pathology detection task and discuss the background and implementation of our proposed method. Section \ref{results} contains the description of our evaluation task, the results obtained by our methods their comparisons with baseline system implementations. Finally, in Section \ref{discuss} \& Section \ref{conclusion}, we discuss the limitations of our method, future directions and summarize our findings.




\begin{figure}[]
\includegraphics[width=\linewidth]{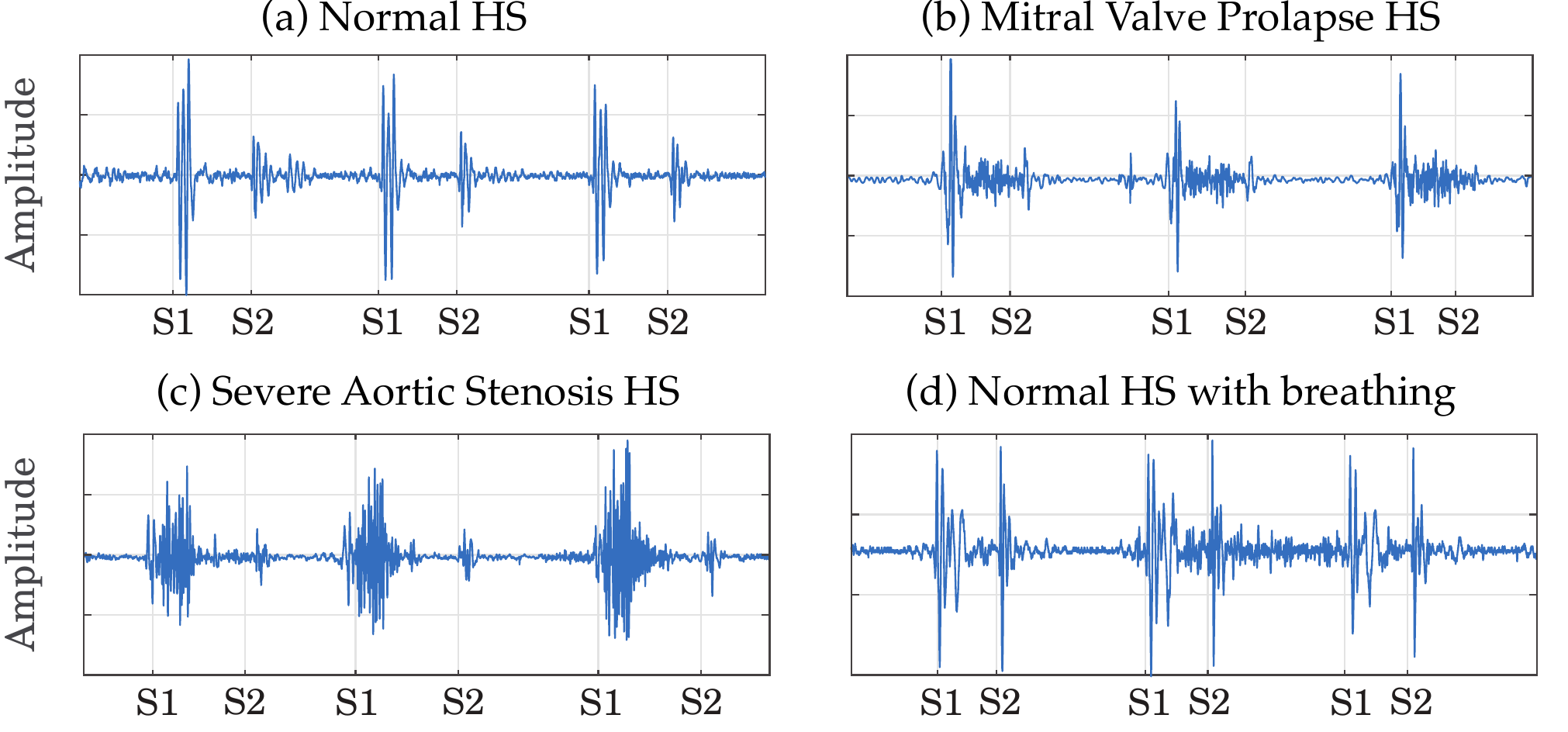}
\centering
\caption{Waveform of different PCG recordings with S1 and S2 \emph{HS} annotations showing (a) Normal \emph{HS}, (b) \emph{HS} with systolic uniform shaped murmur due to Mitral Valve Prolapse, (c) PCG of heart with severely stenosed aortic valve (The murmur has a crescendo-decrescendo shape with almost non-distinguishable S1), and (d) Normal \emph{HS} with breathing sound components similar to a diastolic murmur.
}
\label{shapeHeart}
\end{figure}

\begin{figure}[]
\includegraphics[width=\linewidth]{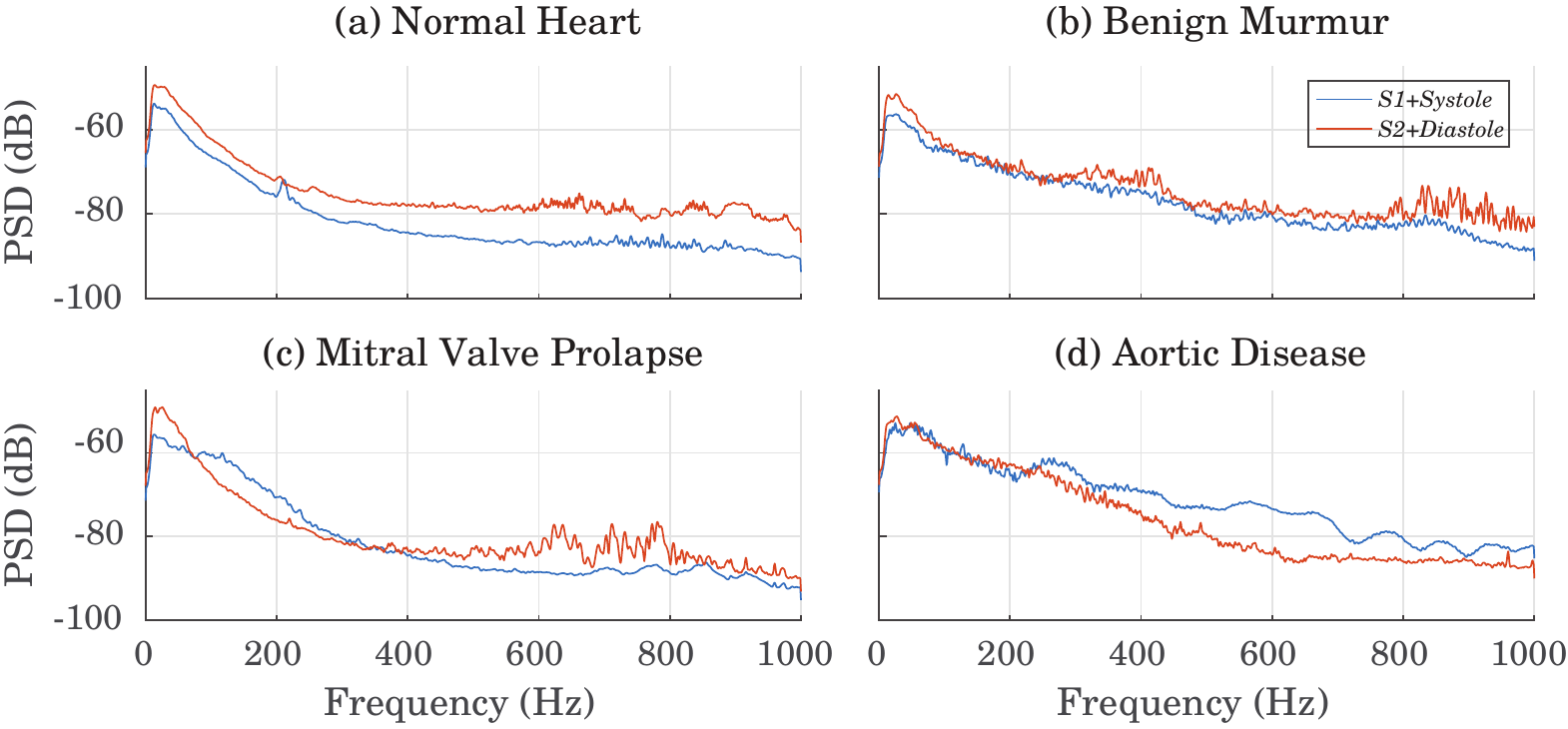}
\centering
\caption{Spectral Average (SA) over Systole and Diastole cardiac phases for different classes of \emph{HS} recordings from the PHSDB subset-$a$ (see Section \ref{dataset}). Spectrum is calculated using the Welch method with a Kaiser window of length $1024$ \& $\beta = 5$. Sub-figure (a) depicts SA of a Normal \emph{HS} recording while Sub-figure (b-d) show SA of Abnormal \emph{HS} Recordings. (b) Recordings with benign murmurs show indication of higher power spectral density (PSD) between $300$-$500$ Hz compared to (a). (c) Mitral Valve Prolapse exhibits higher PSD between $100$-$200$ Hz in Systole. (d) Recording with Aortic Valve Disease have higher energy density in Systole than in Diastole, which is characteristic to Aortic Stenosis.}
\label{sysdia}
\vspace{-3mm}
\end{figure}

\section{Background}
In this section, we provide an overview of heart sounds in general and discuss the origin of abnormal cardiac sounds, e.g., ``murmurs". The importance of the phase of the ``murmurs" with respect to the cardiac cycle is discussed, as it is relevant to the methods proposed in this work. We then discuss possible noise sources in stethoscope recordings and their spectro-temporal properties relating to heart sounds.   
\label{pcg}
\subsection{The First \& Second Heart Sounds}
The electrical activity of the heart can be transduced at the surface of the chest as electrocardiography (ECG) while the mechanical activity can be heard as sounds emanating from the cardiovascular system, which are termed as heart sounds, or the PCG (Fig. \ref{shapeHeart}).
Each PCG recording consists of an underlying periodic structure originating from the cardiac activity of a human heart.

The sounds that a human heart makes in each cardiac cycle, are generally identified as the first ($S1$), second ($S2$), third ($S3$) \& fourth ($S4$) \emph{HS} depending on their temporal occurrence within the cardiac cycle \cite{healioweb}.
The sounds are generally made by the heart valves that control the direction of blood-flow \cite{peskin1972flow}. The first \emph{HS} ($S1$) is generated by the mitral \& tricuspid valves, while the second \emph{HS} ($S2$) is created by the aortic \& pulmonic valves \cite{reddy1985third}. The third ($S3$) and fourth ($S4$) heart sounds are audible in early and late diastole which may indicate congestive heart failure for adults \cite{reddy1985third}.  
The $S1$ and $S2$ sounds mark the onset of the systole and diastole cardiac cycles, respectively. 


\subsection{Cardiac Murmurs}\label{murmurs}
Certain components symptomatic of various cardiac disorders can be present in the PCG signal. These additional audible components in PCG caused due to turbulent blood-flow, are known as ``murmurs" \cite{healioweb}. 
The most common cause of murmurs is diseases in the cardiac structures, especially heart valves. Damaged valves can fail to open and get constricted (termed valvular stenosis) or can fail to close completely, causing leakage during an increase in differential pressure (termed valvular insufficiency). Besides these causes, congenital anomalies, i.e. an orifice in the heart septum, can also cause turbulent blood flow.


The timing and shape of the murmur are crucial for an accurate diagnosis. A murmur can occur during either phase of the cardiac cycle or it can be continuous throughout, exhibiting a change in the power spectral density of the cardiac phase compared to healthy HS (Fig. \ref{sysdia}). The shape of a murmur can be either crescendo (rising), decrescendo (falling), crescendo-decrescendo or uniform (Fig. \ref{shapeHeart}). Stenosis in the aortic or pulmonary valves results in systolic, crescendo-decrescendo murmurs (Fig. \ref{shapeHeart}-c). Mitral and tricuspid regurgitation, on the other hand, are uniformly present throughout systole, at times overwhelming the $S1$ and $S2$ sounds (Fig. \ref{shapeHeart}-b). Aortic and pulmonary regurgitation are diastolic murmurs, both exhibiting a decrescendo envelope. Mitral stenosis also occurs during diastole, with a unique decrescendo-crescendo shape \cite{healioweb}. Clinicians with a trained ear can accurately characterize an audible murmur for differential diagnosis. 

\subsection{Noise in PCG Signals}
PCG is highly prone to noise which tends to be an obstacle to an accurate diagnosis. The noise introduced into a recording can be from internal or external interference. 
Respiration, swallowing, coughing, bowel movement, etc. cause internal interference while environment sounds, vehicle sounds, conversational speech, etc. are considered external \cite{nunes2015noise}. Noise can also assume the tonal characteristics of a murmur, possibly resulting in false positives. For example, in Fig. \ref{shapeHeart}-d, breathing noise in a normal \emph{HS} recording bears the shape of a diastolic murmur, which can confuse cardiac screening systems. Noise related to other pathologies e.g. low pitched wheeze and bronchial breathing can be confusing even for trained cardiologists \cite{obaidat1993phonocardiogram}.

\section{Available Datasets}
\label{dataset}
\subsection{2016 PhysioNet Heart Sound Database (PHSDB)}
The 2016 PhysioNet/CinC Challenge Database \cite{liu2016open} is a cross-corpus archive of PCG recordings from seven different research groups, with each data subset denoted by the letters-\{$a$-$g$,$i$\}. Of the eight different subsets available, data from subsets-\{$a$-$f$\} are compiled in the a publicly available training set and subsets-\{$b$-$e$,$g$,$i$\} are compiled in test set which is not publicly accessible. The training set contains $3153$ \emph{HS} recordings collected from $764$ patients with a total number of $84,425$ cardiac cycles ranging from $35$ to $159$ bpm. The dataset has an unbalanced distribution between classes, with $2488$ \emph{Normal} and $665$ \emph{Abnormal} \emph{HS} recordings. Meta-data on recording quality is provided for all of the recordings, while pathology annotations are available for subsets-\{$a$-$c$,$e$\} (see Table \ref{tableDist}). Cardiac cycle annotations for the onset of $S1$, $S2$, \emph{Systole} \& \emph{Diastole} are available for good quality PCGs. In our experiments, we employ the publicly available training set of this dataset, which from here on is referred to as \emph{PHSDB}.

\subsection{2018 INTERSPEECH ComParE Dataset (HSSDB)} \label{HSS dataset}
The INTERSPEECH 2018 Computational Paralinguistics (ComParE) Challenge \cite{schuller2018interspeech} released the Heart Sounds Shenzhen Database (HSSDB) containing $845$ recordings from $170$ different subjects. The recordings were collected from patients with coronary heart disease, arrhythmia, valvular heart disease, congenital heart disease, etc., without any pathology meta-data. 
The PCG recordings are sampled at $4$ KHz and annotated with three class labels: (i) \emph{Normal}, (ii) \emph{Mild Abnormality}, and (iii) \emph{Moderate/Severe Abnormality}. Class annotations are available for the $502$ training and $180$ development set recordings. The dataset is referred as \emph{HSSDB} from here on. 

\section{Challenges in the Datasets} \label{chall}
\subsection{Heterogeneous Distribution of Samples}\label{heteroDist}
One of the key challenges for the task is the non-homogeneous distribution of pathologies among different subsets. Table \ref{tableDist} contains the number of recordings, subjects and metadata information made available with each of the subsets. Considering both the PHSDB and the HSSDB, we have a total of $2599$ Normal recordings and $1199$ Abnormal recordings from $1087$ different patients. $71.8\%$ of the Normal Recordings were acquired at the Biomedical Engineering Laboratory in DLUT, China \cite{liu2016open} using MLT201 Electret Stethoscope or a piezoelectric sensor-based stethoscope, which was not specified separately for each recording. On the contrary, most of the Abnormal recordings were collected from uncontrolled environments, i.e. hospitals or in-house visits. Due to the lack of homogeneity in the data, it cannot be reliably inferred whether the extracted feature representations are truly discriminatory between pathologies or not \cite{kay2017}. 

\begin{table*}[t]
\centering
\caption{Distribution of different class and pathologies within each subset. Conditions include: Mitral Valve Prolapse (MVP), Benign, Aortic Disease (AD), Coronary Artery Disease (CAD), Mitral Regurgitation (MR), Aortic Regurgitation (AR).}
\label{tableDist}
\resizebox{\linewidth}{!}{%
\begin{tabular}{|c|c|c|c|c|c|c|}
\hline
Dataset (Subset) & Environment & \begin{tabular}[c]{@{}c@{}}No. \\ Subjects\end{tabular} & \begin{tabular}[c]{@{}c@{}}Normal \\ Recordings\end{tabular} & \begin{tabular}[c]{@{}c@{}}Abnormal \\ Recordings\end{tabular} & \begin{tabular}[c]{@{}c@{}}Pathology Annotations \\ (Count)\end{tabular} & Acquisition Device \\ \hline
MITHSDB (a) & \begin{tabular}[c]{@{}c@{}}In-house visit\\ + Hospital\end{tabular} & 121 & 117 & 292 & \begin{tabular}[c]{@{}c@{}}MVP (134), Benign (118),\\ AD (17), MPC (23)\end{tabular} & Welch Allyn Meditron \\ \hline
AADHSDB (b) & Hospital & 106 & 385 & 104 & CAD (104) & 3M Littmann E4000 \\ \hline
AUTHHSDB (c) & Hospital & 31 & 7 & 24 & MR (12), AS (12) & AUDIOSCOPE \\ \hline
UHAHSDB (d) & Hospital + Lab & 38 & 27 & 28 & Not specified & Infral Corp. Prototype \\ \hline
DLUTHSDB (e) Abnormal & Hospital & 335$^*$ & 0 & 151 & CAD (151) & 3M Littmann \\ \hline
DLUTHSDB (e) Normal & Lab & 174 & 1867 & 0 & None & MLT201/Piezo \\ \hline
SUAHSDB (f) & Hospital & 112 & 80 & 34 & Not specified & JABES \\ \hline
HSSDB & Hospital & 170 & 116 & 566 & Not specified & Eko CORE Bluetooth \\ \hline
\multicolumn{7}{l}{\footnotesize $^*$Some subjects in DLUTHSDB (e) Abnormal contained data errors. We only used 151 usable abnormal recordings from this set.}
\end{tabular}
}
\end{table*}
\subsection{Domain Variability Analysis}
To systematically quantify the domain-wise dependency of PCG recordings from different subsets of PHSDB and HSSDB, we design two separate qualitative and quantitative experiments. 
We extract the $6373$ dimensional hand-engineered audio feature set defined in the INTERSPEECH 2018 ComParE Challenge \cite{schuller2018interspeech} as a standard\footnote{The feature extraction codes are available in the openSMILE module \cite{openSMILE} and details of the extracted features can be found in \cite{compare1st}.}. 

The qualitative experiment incorporates visualizing two dimensional t-Stochastic Neighbour Embedding (t-SNE) \cite{TSNE} of the ComParE features. In the quantitative experiment, we examine supervised and unsupervised domain classification performance.
Assuming that the sensor variability accounts for most of the inter-domain variance, we consider the stethoscope type as ground truth for this task. Both \emph{Normal} and \emph{Abnormal} HS recordings are included in this experiment to monitor the effect of class-wise variance within the domains. We use a Decision Tree classifier and a Kmeans {(with $k=8$)} clustering algorithm for the supervised and unsupervised experiments, respectively. In a $4$-fold cross-validation task, the Decision Tree was able to classify between acquisition devices with $95.1\%$ accuracy; whereas the Kmeans algorithm obtained an Adjusted Random Index Score (similarity measure) of $63.7\%$ and Completeness Score (accuracy measure for intra-class assignment) of $68.8\%$. Fig. \ref{tsne}-a depicts the t-SNE embeddings of the $6,373$ dimensional feature space with stethoscope annotations for qualitative evaluation, where distinct manifolds for each sensor can be seen. Significant overlap is also observed during a qualitative comparison between the predicted clusters by Kmeans and the sensor annotations. Kmeans forms two separate clusters for \emph{Normal HS} from data subset DLUTHSDB (subset-e), which is most likely due to the fact that it was recorded using two different stethoscopes (MLT201/Piezo) \cite{liu2016open}. Finally, we note that none of the underlying manifolds significantly correspond to class differences, indicating that the feature variance due to the sensor/domain is predominant compared to the cardiac condition. Thus domain mismatch is of significant concern in the HS classification task.


\section{Proposed Method}\label{method}
In this section, we discuss the proposed FIR-like CNN layers used as the front-end to our deep learning system, to filter out patterns specific to domain differences via loss optimization.
\subsection{Background and Overview}
\label{tconv}
\subsubsection{CNNs are analogous to FIR filters}
Let, a CNN Architecture of $L$ convolutional layers has kernel weights $\vec{h}_{c,k}^{l} = [h_{c,k}^{l}(0),h_{c,k}^{l}(1),...,h_{c,k}^{l}(K^{l}-1)]$ 
where the superscript $l=0,1,...,L-1$ denotes the layer index, subscript $c$ \& $k$ represent the input channel \& kernel indices, and $K^{l}$ is the kernel size of the $l$-th layer respectively. Vector $\vec{x}_{c}^{l} = [x_{c}^{l}(0),x_{c}^{l}(1),...,x_{c}^{l}(N-1)]$ represents the $c$-th channel, $N$ sample input to the $l$-th layer, while $\vec{z}_{k}^{l} = [z_{k}^{l}(0),z_{k}^{l}(1),...,z_{k}^{l}(N-1)]$ denotes the activation from the $k$-th kernel of the $l$-th CNN layer. We assume that each layer has a linear activation, have odd length kernels and zero bias; such 1DCNN units are termed as \emph{tConv} \cite{sainath2015google}. For a $C$ channel input, the operation of a \emph{tConv} kernel can thus be expressed as a weighted average of the input given by:
\begin{eqnarray}\label{eq1}
z_{k}^{l}(n) = \sum_{c=0}^{C} \sum_{i=0}^{K^{l}-1}h_{c,k}^{l}(i) \hspace{0.8mm} x_{c}^{l}\Big(n + \frac{K^{l}-1}{2} - i\Big) \\
\vec{x}_{c}^{l} = \vec{z}_{k}^{l-1} \hspace{1.8mm} \text{for $l=1,2,...,L-1$}. \nonumber
\end{eqnarray}
For a single channel input on the first layer of a CNN, (\ref{eq1}) can be simplified as:
\begin{equation}\label{eqtConv}
z_{k}^{0}(n) = \sum_{i=0}^{K^{0}-1}h_{0,k}^{0}(i) \hspace{0.8mm} x_{0}^{0}\Big(n + \frac{K^{0}-1}{2} - i\Big) .
\end{equation}
This is analogous to the operation of a $K$+$1$ order FIR filter with coefficients $\vec{h} = [h(0),h(1),...,h(K-1)]$ given by:
\begin{eqnarray}
y(n) &=& \sum_{i=0}^{K-1}h(i)x(n-i)
\end{eqnarray}
where, $\Vec{x} = [x(0),x(1),...,x(N-1)]$ is a single channel input to the FIR filter and $\Vec{y} = [y(0),y(1),...,y(N-1)]$ is the output. Therefore, for a single channel vector, each kernel from the first layer of a CNN architecture acts as an FIR filter and performs time-frequency decomposition with coefficient vector $\vec{h}_{c,k}^{l}$ of shape $1 \times K^{l}$. For consecutive layers, the operation is similar to cross-correlation where the shape of the kernel is $C^{l} \times K^{l}$, where $C^{l}$ is the number of input channels to that layer. Theoretically, the first-layer of a CNN, should learn the optimal time-frequency decomposition of the input signal by stochastic gradient descent. In practice, the optimization process is susceptible to the kernel size, number of kernels in the front-end and the volume of training data.

\subsubsection{Symmetry condition for a causal generalized linear phase FIR filter}
A generalized linear phase system $h(n) \xrightarrow{\mathcal{F}} H(e^{jw})$ is one that has a constant group delay, i.e. its phase angle can be expressed as:
\begin{gather}\label{eq2}
    \angle H(e^{jw}) = -aw+ B \text{\hspace{5mm}for, } 0<w<\pi \\
    grd\{H(e^{jw})\} = -\frac{d}{dw}\{\angle H(e^{jw})\} = a \nonumber
\end{gather}
where, $a$ and $B$ are real coefficients, $w$ is the normalized frequency, $\mathcal{F}$ denotes the Fourier transform operation, and $grd\{.\}$ is the group delay. Taking the tangent of the phase angle from (\ref{eq2}) and using the Euler's formula on the Fourier transform of $h(n)$, we obtain the following condition for a causal generalized linear phase system:
\begin{equation}\label{eq3}
    \sum_{n=0}^{\infty}h(n)\sin(w(n-a)+B)=0.
\end{equation}
For a causal system with FIR length $K+1$, (\ref{eq3}) is satisfied when:
\begin{gather}
    2a = K = \text{an integer}\nonumber\\
    \label{eq4}
    h(K-n) = h(n) \text{ or, } h(K-n) = -h(n)
\end{gather}
This implies that an FIR filter with integer length has linear phase response when its coefficients are symmetric/anti-symmetric \cite{oppenheim1999discrete}. The theory can similarly be extended to \emph{tConv} units to ensure linear phase (LP) response.

\subsection{Implementation of tConv variants}
The \emph{tConv} units in the front-end enable the pre-processing steps, e.g. spectral decomposition or filterbank analysis, to be supplemented by the first layer of an end-to-end CNN. This can eradicate the need for empirical tuning of the pre-processing pipeline. 
However, it adds another stage of complexity to the optimization process. In Section \ref{murmurs}, we discussed that diagnosis of PCG abnormality is subjected to the shape (Fig. \ref{shapeHeart}) and phase of cardiac murmurs withing the cardiac cycle. Thus, the front-end kernels should have a linear phase response to retain the shape of the PCG. 
\subsubsection{Linear Phase tConv}
According to (\ref{eq4}), a kernel with integer length and symmetric weights around its center would have linear phase, i.e. it would introduce an equal delay for all of the passing frequencies/patterns, ensuring no distortion. Depending on the type of symmetry and the parity of the kernel length, Linear Phase \emph{tConvs} can be of four different types as in Table \ref{typeTable}.
\begin{table}[t]
\centering
\caption{Types of Linear Phase \emph{tConv} layers and their kernel properties}
\label{typeTable}
\begin{tabular}{@{}ccc@{}}
\toprule
\emph{tConv} Type & Kernel Shape & Length \\ \midrule
Type I & \multirow{2}{*}{Symmetric} & Odd \\
Type II &  & Even \\ \midrule
Type III & \multirow{2}{*}{Anti-symmetric} & Odd \\
Type IV &  & Even \\ \bottomrule
\end{tabular}
\end{table}
We implement the LP-\emph{tConv} units using CNN kernels with weight sharing between coefficients on the two sides of the symmetry. This results in half of the kernel weights being learned and shared.
\subsubsection{Zero Phase tConv}
A filter $h(n) \xrightarrow{\mathcal{F}} H(e^{jw})$ is termed a zero phase filter when $grd\{H(e^{jw})\} = 0$  and  $\angle H(e^{jw}) = 1$. Incorporating a forward-reverse convolution \cite{shi2006ZPtConv}, we propose a zero phase \emph{tConv} layer that has no phase effect on the input signal. If $x(n)$ is the input signal, $h(n)$ is the impulse response of the kernel, and $y(n)$ is the output, we have in the frequency domain:
\begin{equation}
\begin{split}
Y(e^{jw})&=X(e^{jw}).H(e^{jw}).H^{*}(e^{jw})\\
&=X(e^{jw})|H(e^{jw})|^{2}
\end{split}
\end{equation}
where, $h(-n) \xrightarrow{\mathcal{F}} H^{*}(e^{jw})$. Note that, the flip operation in time domain is equivalent to taking the complex conjugate in the frequency domain. Therefore, the effect of a ZP-\emph{tConv} is a simple multiplication by the squared magnitude in the frequency domain. In out implementation of the ZP-\emph{tConv} unit, we perform two consecutive convolution operations with the same kernel; during the second convolution, we flip the kernel to equalize the phase response of the first convolution.
\subsubsection{Gammatone tConv}
The gammatone auditory filterbank is implemented in practice as a series of parallel band-pass filters; it models the tuning frequency at different points of the human basilar membrane \cite{patterson1987efficient}. 
Since the auscultation of heart sounds incorporates the auditory perception of the cardiologist, we introduce a novel \emph{tConv} unit that approximates a gammatone function; the output of these gammatone \emph{tConvs} would be similar to the features extracted by the basilar membrane. The gammatone impulse response is given by:
\begin{equation}\label{eqgamma}
g(n) = \alpha n^{\eta-1}e^{-2\pi \beta n}\cos(2\pi fn + \phi)
\end{equation}
where, $g(n)$, $\alpha$, $\eta$, $\beta$, $f$ and $\phi$ denote the $n$-th gammatone coefficient, amplitude, filter order, bandwidth, center frequency and phase of the gammatone wavelet (in radians), respectively. 
In our implementation, we set $\phi=0$ and condition the units to learn kernel coefficients that follow (\ref{eqgamma}). Therefore, a gammatone \emph{tConv} has only $4$ learnable parameters ($\alpha$,$\eta$,$\beta$,$f$); the gradients of which can be calculated using backpropagation. Let, $\mathcal{L}$ be the loss being minimized by the CNN, $\vec{x} \in \mathbb{R}^{1\times N}$ a single channel input to the gammatone front-end, $\vec{g_{k}^{0}} \in \mathbb{R}^{1\times K^{0}}$ the $k-th$ kernel, $K^{0}$ the kernel length, and $\vec{z_{k}^{0}}$ its activation. Backpropagation involves calculating gradients for each of the four learnable parameters of the $k$-th gammatone \emph{tConv}:
\begin{gather}
    \frac{\partial\mathcal{L}}{p_{k}^{0}} = 
    \sum_{i=0}^{K^{0}-1} 
    \frac{\partial\mathcal{L}}{\partial g_{k}^{0}(i)} \frac{\partial g_{k}^{0}(i)}{\partial p_{k}^{0}} \text{ for, } p \in \{\alpha,\eta,\beta,f\}.\nonumber
\end{gather}
Using (\ref{eqtConv}), we can show that the loss gradient for the $i$-th gammatone kernel coefficient is:
\begin{align}
    \frac{\partial\mathcal{L}}{\partial g_{k}^{0}(i)} &= \sum_{n=0}^{N-1} 
    \frac{\partial\mathcal{L}}{\partial z_{k}^{0}(n)} \frac{\partial z_{k}^{0}(n)}{\partial g_{k}^{0}(i)}\nonumber\\
    &= \sum_{n=0}^{N-1} 
    \frac{\partial\mathcal{L}}{\partial z_{k}^{0}(n)} x \left(n+\frac{K^{0}-1}{2}-i\right).\nonumber
\end{align}
Therefore,
\begin{equation}
    \frac{\partial\mathcal{L}}{p_{k}^{0}} = 
    \sum_{i=0}^{K^{0}-1} \frac{\partial g_{k}^{0}(i)}{\partial p_{k}^{0}} \sum_{n=0}^{N-1} 
    \frac{\partial\mathcal{L}}{\partial z_{k}^{0}(n)} x\left(n+\frac{K^{0}-1}{2}-i\right)
\end{equation}
where, $\frac{\partial\mathcal{L}}{\partial z_{k}^{0}(n)}$ is derived from backpropagation on the following layers.
\subsection{Proposed Model Architecture}
\label{model}
We have employed a branched CNN architecture \cite{potes2016ensemble} for end-to-end pathology detection from cardiac cycles. The front-end of the model is a learnable filterbank, built with four \emph{tConv} units. Each of the spectral bands decomposed by the learnable filterbank is passed through a separate branch of our CNN architecture. Each branch has two convolutional layers of kernel size $5$, followed by a Rectified Linear Unit (ReLU) activation and a max-pooling of $2$. Activations are normalized for each training mini-batch prior to ReLU and dropped out with a probability of $.5$ as regularization in each branch. The first convolutional layer has $8$ filters while the second has $4$. The outputs of the four branches are fed to an MLP network after being concatenated along the channels and flattened. The MLP network has a hidden layer of $20$ neurons with ReLU activation and two neurons as outputs with softmax activation. Cardiac cycles are extracted from each PCG resampled to $1$kHz using the method presented in \cite{springer2016segmentation} and zero-padded to be $2.5$s in length before being fed into our branched CNN architecture. The posterior predictions for all of the cardiac cycles are fused for each recording. 
Cross-entropy loss is optimized using the Adam optimizer. The learning rate and hyperparameters are set according to \cite{humayunFIR} via $4$-fold cross-validation. Codes are available at \texttt{Github}\footnote{Implementation details at \textbf{\texttt{\href{https://github.com/AhmedImtiazPrio/heartnet}{github.com/mHealthBuet/heartnet}}}}.


\begin{figure}[]
\includegraphics[width=\linewidth]{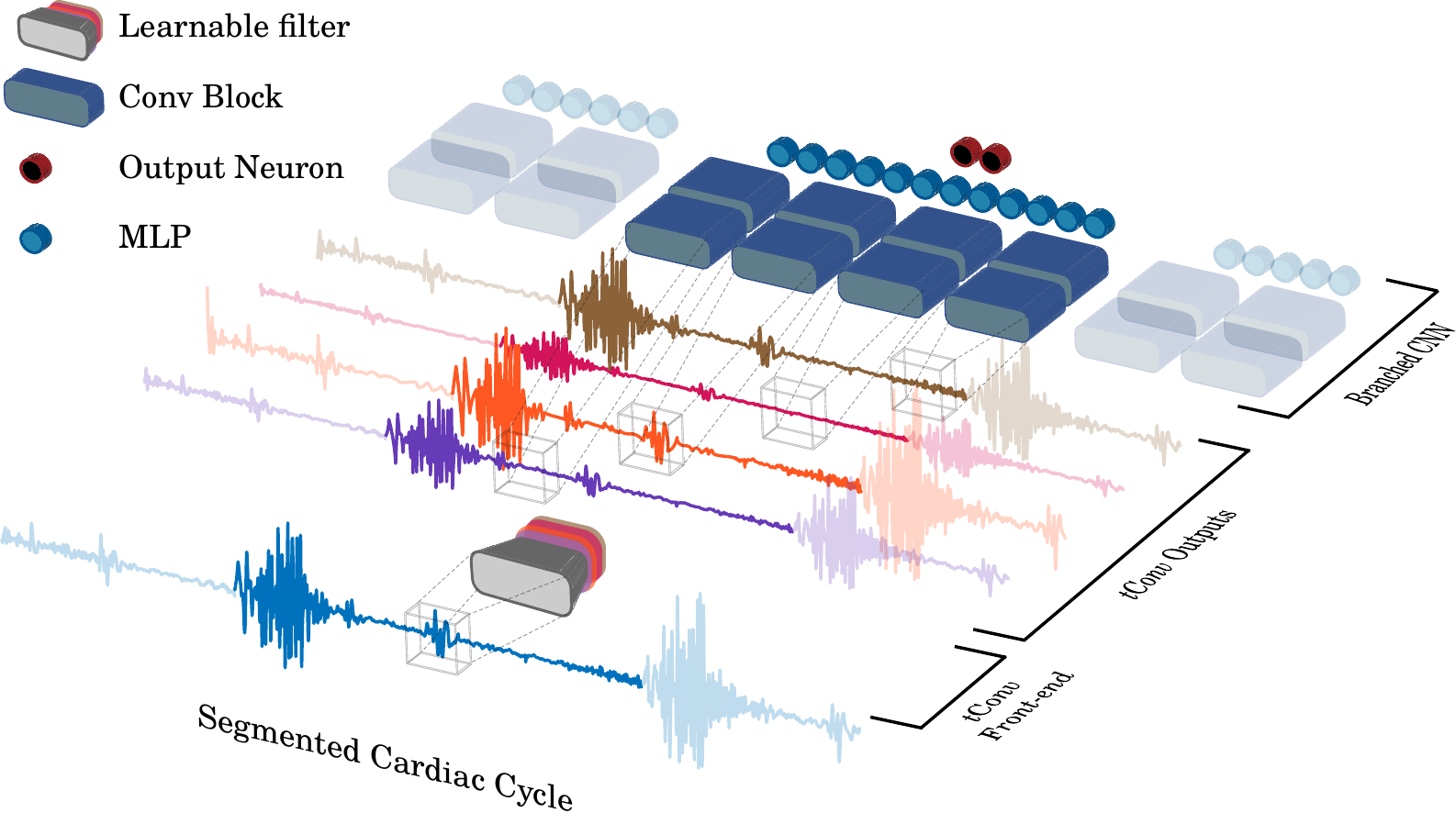}
\centering
\caption{Proposed branched CNN architecture including a \emph{learnable filterbank} front-end comprising of \emph{tConv} units. {Each convolutional block in the figure consists of a convolution layer, batch normalization layer, ReLU activation, Dropout and a pooling layer.} 
}
\label{propmod}
\end{figure}

\subsection{Proposed Training Regime: Domain Balanced Training}
To account for the non-homogeneous distribution of data, we train our deep learning model with each mini-batch of size $B$, balanced with an equal number of classes from each PHSDB subset. We refer to this method as Domain Balanced Training (DBT). During DBT, $12$ separate queues are formed for each class from the six different PHSDB subsets in training. An equal number of samples from each queue is drawn sequentially to populate each training mini-batch. The effective mini-batch size ($B_{\mbox{\scriptsize eff}}$) is therefore:
\begin{equation}
    B_{\mbox{\scriptsize eff}} = 12*\lfloor B/12 \rfloor \nonumber
\end{equation}
where $\lfloor . \rfloor$ is the floor operation. Each queue is shuffled and reinitialized asynchronously if all of the cardiac cycles within it is drawn. Each training epoch contains a specified number of mini-batch iterations. We train the proposed methods with $B=64$, for $300$ epochs with mini-batch iterations per epoch equal to the number of subset-$a$ cardiac cycles in training divided by $B_{\mbox{\scriptsize eff}}$. 

\begin{figure}[]
\includegraphics[width=\linewidth]{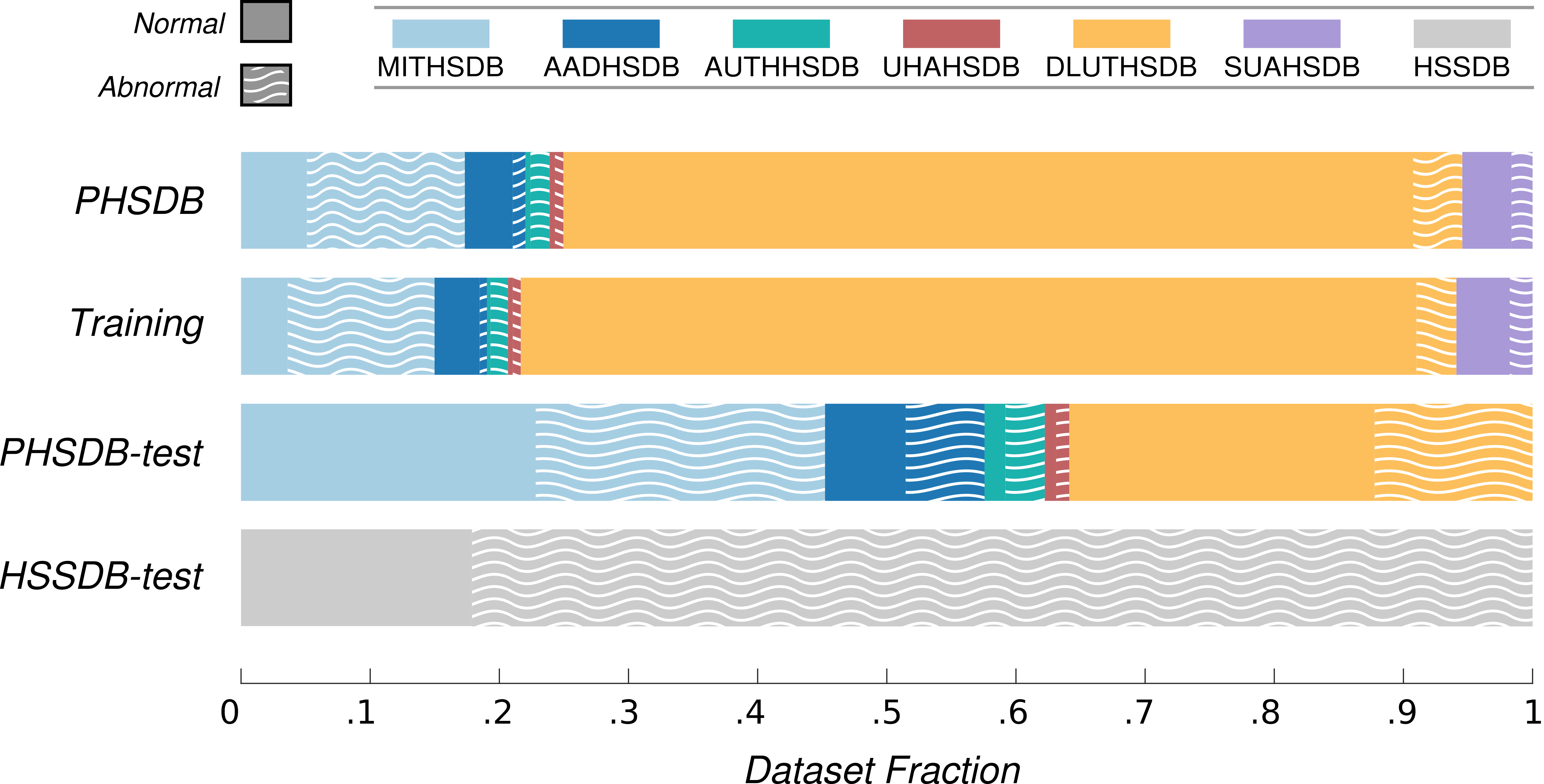}
\centering
\caption{Distribution of cardiac cycles from each subset of PHSDB \& HSSDB in the training and evaluation sets. PHSDB-test consists of an equal number of \emph{Normal} and \emph{Abnormal} recordings.}
\label{foldsplit}
\vspace{3mm}
\end{figure}

\section{Experiments and Results}
\label{results}
\subsection{Evaluation Setup and Metrics}\label{evalSetup}
We design two different evaluation setups from the data available from PHSDB and HSSDB. While data from only PHSDB is used for training, $301$ multi-domain PCG recordings from PHSDB are held-out for evaluation which is referred to as \emph{PHSDB-test}.
This set was made publicly available as the validation set for the Physionet 2016 CinC Challenge \cite{liu2016open}. The second testing setup is comprised of $682$ recordings from HSSDB from here on referred to as \emph{HSSDB-test}. For both cases, the training set contains $2815$ PCG recordings from PHSDB. Fig. \ref{foldsplit} shows the distribution of \emph{Normal} and \emph{Abnormal} cardiac cycles from each subset of PHSDB and HSSDB in our evaluation setups. We note that the HSSDB dataset is never used for model training and is only used for testing.
The metrics calculated to measure overall performance are Sensitivity, Specificity, F1-score and modified accuracy (Macc) (average of sensitivity and specificity). During experiments on PHSDB-test, we also monitor the accuracy for each subset and their average to evaluate the performance of the methods across different domains that are seen during training. On the other hand, experiments on HSSDB-test assess the performance of the methods on previously unseen distribution.

\begin{table}[]
\centering
\caption{Comparison between methods proposed by Gabor et al. \cite{Gabor2018} on \emph{PHSDB-test}}
\label{baselineVal}
\resizebox{\linewidth}{!}{%
\begin{tabular}{@{}ccccccc@{}}
\toprule
Sampling/No.SVC & Features & Avg Subset Acc. & Sens. & Spec. & Macc & F1 \\ \midrule
\multirow{4}{*}{\begin{tabular}[c]{@{}c@{}}Downsampling\\ 100 SVC\end{tabular}} & BoAW & 63.84 & \textbf{79.47} & 56.84 & 68.15 & \textbf{71.85} \\
 & FBANK & 65.18 & 75.49 & 58.21 & 66.85 & 69.93 \\
 & ComParE & 61.92 & 67.54 & 63.01 & 65.28 & 66.45 \\
 & \emph{Fusion} & 63.43 & 78.14 & 56.84 & 67.49 & 71.08 \\ \midrule
\multirow{4}{*}{\begin{tabular}[c]{@{}c@{}}Upsampling\\ 5 SVC\end{tabular}} & BoAW & 64.66 & 72.18 & 67.12 & \textbf{69.65} & 70.78 \\
 & FBANK & \textbf{65.38} & 68.21 & 66.43 & 67.32 & 67.99 \\
 & ComPare & 62.53 & 65.56 & 67.12 & 66.34 & 66.44 \\
 & \emph{Fusion} & 62.45 & 68.87 & \textbf{69.17} & 69.02 & 69.33 \\ \bottomrule
\end{tabular}%
}
\end{table}
\subsection{Baselines Methods and Implementation}\label{baselines}
We use two different baseline models for performance comparison: (i) the top-scoring method from the INTERSPEECH 2018 ComParE Heart Beats Sub-Challenge by Gabor et al. \cite{Gabor2018} as a traditional machine learning baseline, and (ii) the best performing system in the Physionet 2016 CinC Challenge developed by Potes et al. \cite{potes2016ensemble} as a deep learning baseline. 

The Gabor et al. \cite{Gabor2018} implementation employs bagged Support Vector Classification (SVC) ensembles, trained separately with three different feature sets: (i) Bag-of-Audio-Words (BoAW) representation of MFCCs, (ii) $2707$ mel filter bank energy features (FBwhANK), and (iii) $6373$ ComParE features \cite{compare1st}. The BoAW representation is implemented with a codebook size of $4096$  using the OpenXBOW package \cite{schmitt2017openxbow} as specified in \cite{Gabor2018}. Only the training data was used to select the centroids. The training data is balanced in two approaches: undersampling the majority and oversampling the minority class. To account for the randomness while sampling, $100$ SVCs are trained on $100$ undersampled bags of data and $5$ SVCs for $5$ oversampled bags of data. The posterior probabilities from each bagged SVCs are fused for inference from each recording. The regularization parameter (\emph{nu}) for each SVC is set to $0.5$. 

Table \ref{baselineVal} shows PHSDB-test results of the different systems proposed in \cite{Gabor2018}. The \emph{Fusion} results are obtained via a predefined weighted average of the predictions from each ensemble according to \cite{Gabor2018}. The ensemble trained on BoAW features outperform the fusion systems in both the oversampling and undersampling cases. 
The BoAW ensemble and the \emph{Fusion} framework are denoted by Gabor-BoAW-SVC and Gabor-Fusion-SVC, respectively.
The branched CNN model by Potes et al. \cite{potes2016ensemble} (Potes-CNN) is implemented as our deep learning baseline system. It has a static front-end FIR filterbank as the input and provides inferences for each segmented cardiac cycles. The implementation is identical to our proposed architecture except for the front-end \emph{learnable filterbank}. This enables orthogonal comparison between the performance of our proposed methods and the deep learning baseline.

The performance of the baseline systems and proposed methods are evaluated on the class balanced PHSDB-test and unbalanced HSSDB-test sets (Sec. \ref{evalSetup}). PHSDB-test and HSSDB-test results are presented in Tables \ref{valResults} and \ref{testResults}, respectively. 

\subsection{Performance evaluation on PHSDB-test}
In this subsection, we discuss the results on PHSDB-test where {each of the subsets-\emph{\{a-f\}} contain data from a different stethoscope,} as seen in Table \ref{tableDist}. Since the PHSDB-test is sourced from the same distribution as the training data (Fig. \ref{foldsplit}), the models are able to learn from multiple stethoscopes during training. From Table \ref{valResults}, we observe that the proposed methods portray superior performance in all of the metrics compared to the baselines, with a significant improvement in average subset-wise accuracy and Macc. Our proposed CNN with a learnable filterbank front-end with linear phase Type IV \emph{tConv}s, acquired relative improvements of $8\%$ and $11.84\%$ in Macc compared to the Potes-CNN and Gabor-BoAW-SVC (Upsamp.) baselines, respectively. \cite{potes2016ensemble}.
{In both cases, the improvements are found to be significant ($p<0.01$) according to the McNemar's test \cite{dietterich1998approximate}.}


\begin{table*}[]
\centering
\caption{Performance comparison on the PHSDB-test set between baseline and proposed methods}
\label{valResults}
\resizebox{.9\linewidth}{!}{%
\begin{tabular}{@{}ccccccccccc@{}}
\toprule
\multirow{2}{*}{Methods} & \multicolumn{6}{c}{Accuracy in data subsets/domains} & \multirow{2}{*}{Sens.} & \multirow{2}{*}{Spec.} & \multirow{2}{*}{Macc} & \multirow{2}{*}{F1} \\ \cmidrule(lr){2-7}
 & a & b & c & d & e & Avg &  &  &  &  \\ \midrule
\multicolumn{11}{c}{\cellcolor[HTML]{EFEFEF}Baseline Systems} \\
Gabor-BoAW-SVC (Upsamp.) & 50 & 57.14 & 57.14 & 60 & 99.01 & 64.66 & 72.18 & 67.12 & 69.65 & 70.78 \\
Potes-CNN & 60 & 63.26 & 71.43 & 40 & 100 & 66.94 & 82.61 & 64.38 & 73.5 & 74.99 \\
\multicolumn{11}{c}{\cellcolor[HTML]{EFEFEF}Proposed Systems} \\
Potes-CNN DBT & 71.25 & 70.41 & 85.71 & 66.66 & 97.5 & 78.31 & 87.68 & 71.91 & 79.79 & 80.67 \\ 
Type I \emph{tConv}-CNN & 75 & 68.37 & 100 & 60 & 96.62 & 79.99 & 88.41 & 71.23 & 79.82 & 80.79 \\
Type II \emph{tConv}-CNN & 72.5 & 74.49 & 100 & 70 & 94.38 & \textbf{82.27} & \textbf{90.58} & 71.23 & 80.91 & \textbf{81.96} \\
Type III \emph{tConv}-CNN & 72.5 & 64.28 & 71.42 & 80 & 96.63 & 76.97 & 88.42 & 67.1 & 77.64 & 79.22 \\
Type IV \emph{tConv}-CNN & 76.25 & 71.43 & 71.42 & 80 & 97.75 & 79.37 & 86.95 & \textbf{76.02} & \textbf{81.49} & 81.91 \\
ZP \emph{tConv}-CNN & 72.5 & 73.46 & 100 & 60 & 96.62 & 80.52 & 86.96 & 74.65 & 80.81 & 81.35 \\
Gammatone \emph{tConv}-CNN & 75 & 70.41 & 85.71 & 60 & 93.75 & 76.97 & 87.68 & 71.23 & 79.46 & 80.39 \\ \bottomrule
\end{tabular}%
}
\end{table*}

\subsection{Evaluating the Effect of Domain Balanced Training (DBT)}
Our proposed DBT scheme was introduced to tackle the issue of heterogeneous distribution of data as discussed in Section \ref{heteroDist}. During primary investigations with deep learning architectures without DBT, we observed that due to the majority presence of subset-e in the training data, PHSDB-test accuracy for subset-e converged to the maximum, while the accuracy for other subsets remained below par. For both our baselines, Gabor-BoAW-SVC and Potes-CNN, subset-e accuracy maxed out at approximately $99\%$ while the accuracy for other subsets failed to cross $60\%$ (Table \ref{valResults}). This indicates that the models are over-fitted towards the data subset-e while sacrificing performance on other domains.
DBT therefore acts as an adaptation scheme where both the domain and class labels are used to balance the distribution of samples within each training mini-batch. 

The effect of DBT can be observed by comparing the subset-wise accuracy differences between Potes-CNN {(Macc $= 73.5\%$)} and Potes-CNN DBT {(Macc $= 79.79\%$)} in Table \ref{valResults}. Without DBT, PHSDB-test accuracy of subset-e reaches $100\%$ while the accuracy for subsets-{a-d} falls low. This is due to the highly prevalent subset-e data in each training mini-batch when the mini-batches are sampled randomly. With DBT, accuracy for subsets-{a-d} exhibit a median increase of $12.77\%$, while the accuracy for subset-e drops by $2.5\%$. Since each training mini-batch is also class balanced, it increases the sensitivity, specificity and Macc by $5.53\%$, $7.53\%$ and $6.3\%$, respectively.  {The difference between these two systems are significant with $p<0.01$ using McNemar's test, demonstrating the effectiveness of DBT.} Domain differences are further reduced by our proposed linear phase Type II \emph{tConv} learnable filterbank, with an average subset-wise accuracy of $82.27\%$. 



\subsection{Interpreting the learnable filterbanks}
\begin{figure}[t]
\includegraphics[width=\linewidth]{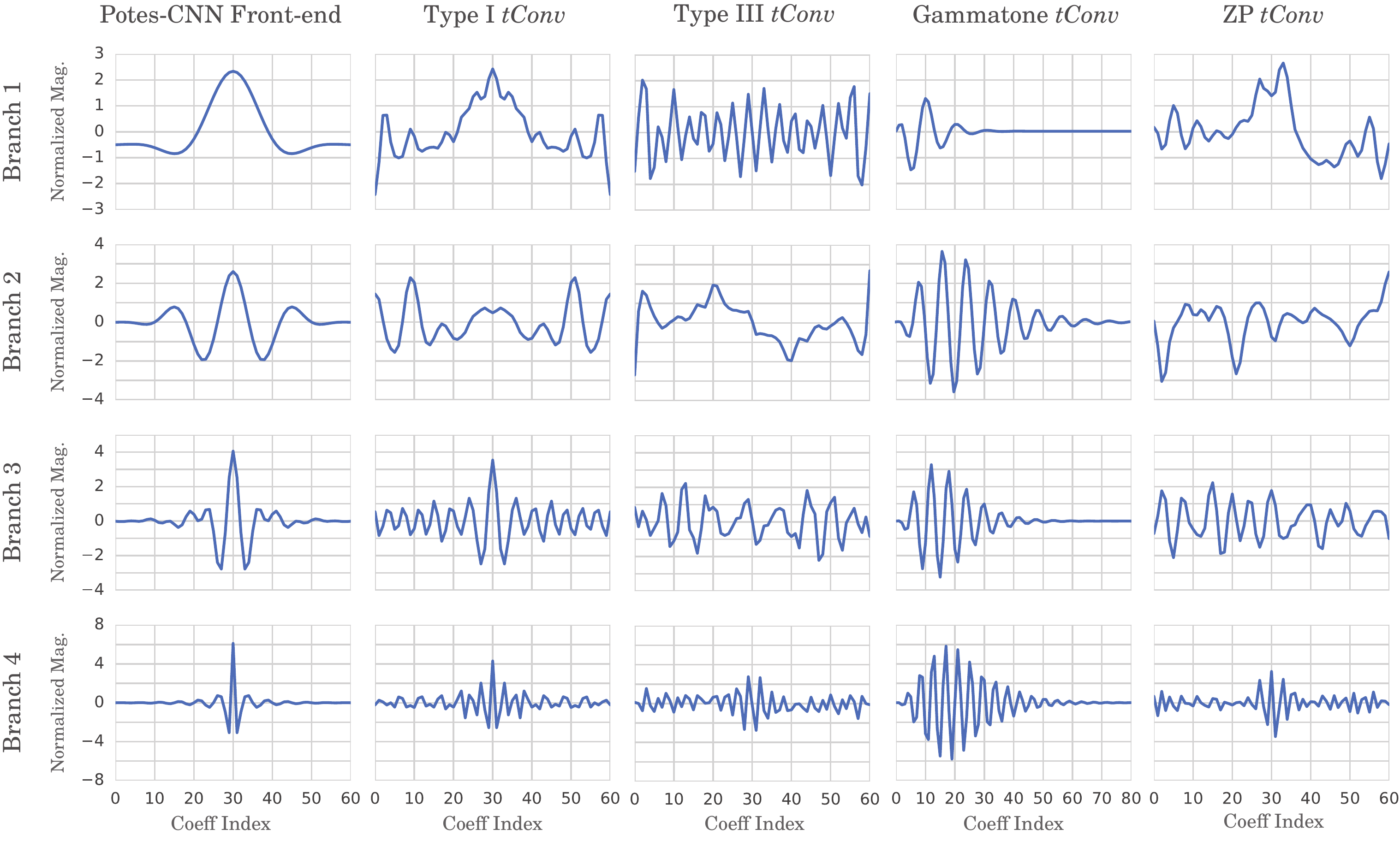}
\centering
\vspace{-2mm}
\caption{{FIR filter} coefficients {(i.e., filter impulse responses)} learned by the front-end kernels (along row) for the baseline static front-end and the proposed \emph{learnable filterbanks} with \emph{tConv} variants (along column).}
\label{coefflearn}
\vspace{5mm}
\includegraphics[width=\linewidth]{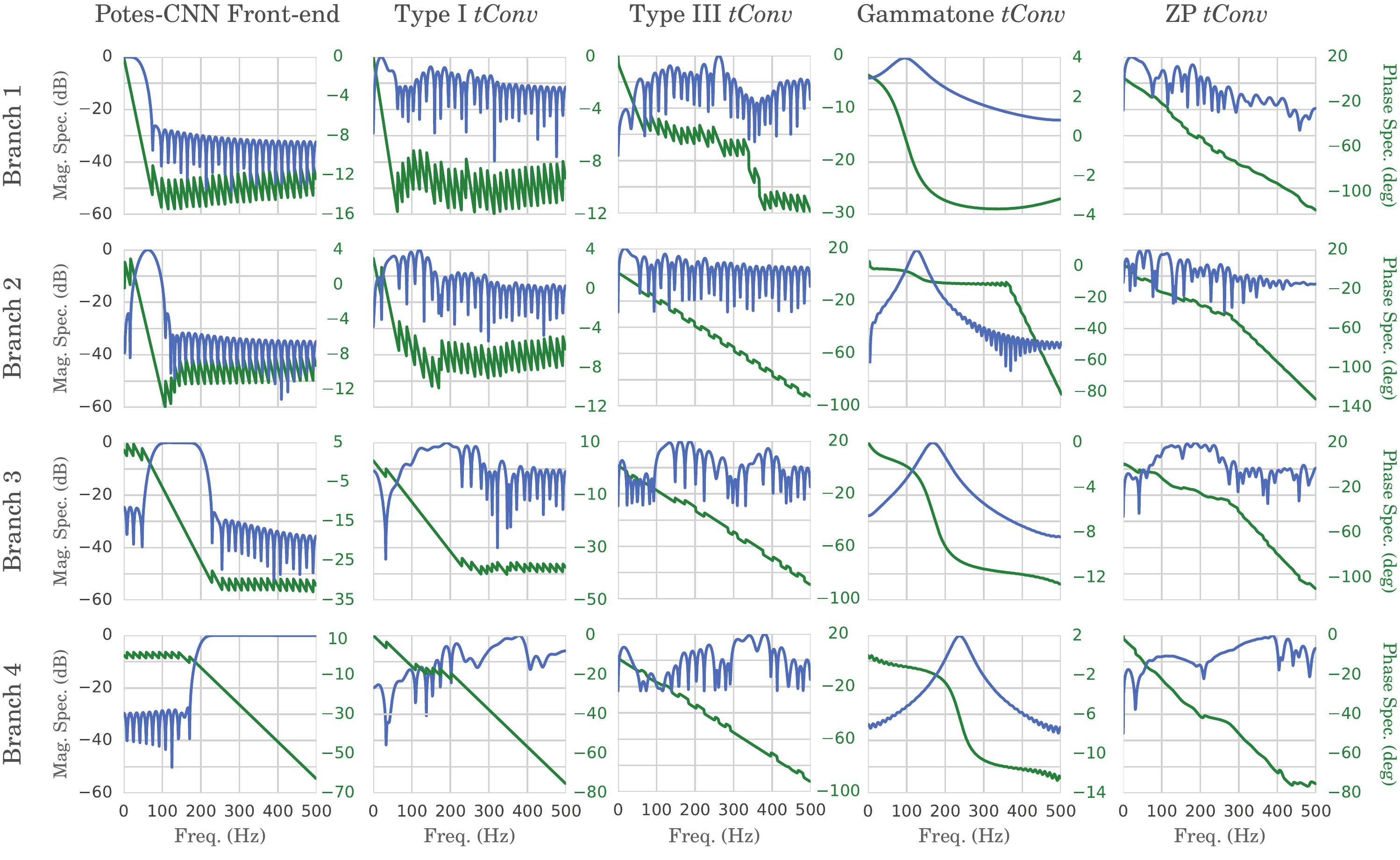}
\centering
\vspace{-4mm}
\caption{Magnitude (Blue) and Phase (Green) spectrum for each kernel in the baseline static front-end and the proposed \emph{learnable filterbanks} with \emph{tConv} variants. Note that the magnitude and phase response of ZP \emph{tConv} in the figure corresponds to its kernel response prior to the bi-convolution operation.}
\label{coefflearnFreq}
\vspace{-3mm}
\end{figure}

An advantage of the \emph{tConv} kernels is that it enables interpretability of the model. As each set of coefficients learned by a kernel is necessarily an impulse response of the particular filter in the learnable filterbank, each kernel can also be interpreted in the Fourier domain. As the coefficients are learned by optimizing the loss specific to our classification task, specific signal properties contributing to each pathology can also be analyzed from these kernels. Fig. \ref{coefflearn} and \ref{coefflearnFreq} show the front-end FIR filter coefficients {(i.e., filter impulse responses)} from different methods and their magnitude/phase responses, respectively. All of the learnable filterbanks are initialized with static FIR coefficients \cite{potes2016ensemble} except for the gammatone \emph{tConv} units. The gammatone \emph{tConv} is initialized with $\alpha_{k}=10^{5}$ and $\eta_{k}=4$, while $f_{k} \sim \mathcal{U}(10,400)$ and $\beta_{k}\sim \mathcal{N}(30,6^{2})$ are randomly sampled from a uniform and a normal distribution,  respectively.

In Fig. \ref{coefflearnFreq}, we observe that the gammatone \emph{tConv} kernels learn to pass center frequencies of $98$, $121$, $182$ and $237$. The filter exhibits a notch magnitude response, with an almost constant group delay in the passband. The Type III \emph{tConv} being an anti-symmetric odd length filter has a zero in the laplacian zero frequency and therefore, fails to form the branch $1$ low-pass filter. The Type I \emph{tConv} forms a low pass and three bandpass filters in the four branches. Most of the front-end kernels have filtered out frequencies above $400$ Hz. In Fig. \ref{sysdia}, we infer that frequencies above $400$ could contain more noise components compared to patterns correlated with pathology. In the case of the ZP-\emph{tConv} layer, the filters tend to assume a linear phase response, even though subsequent operations within the layer equalize the phase effect of the filters. This consolidates that indeed, the optimum front-end kernel for the task would require to have linear/zero phase effect to ensure a distortionless transmission to the activations.

\begin{figure}
\includegraphics[width=\linewidth]{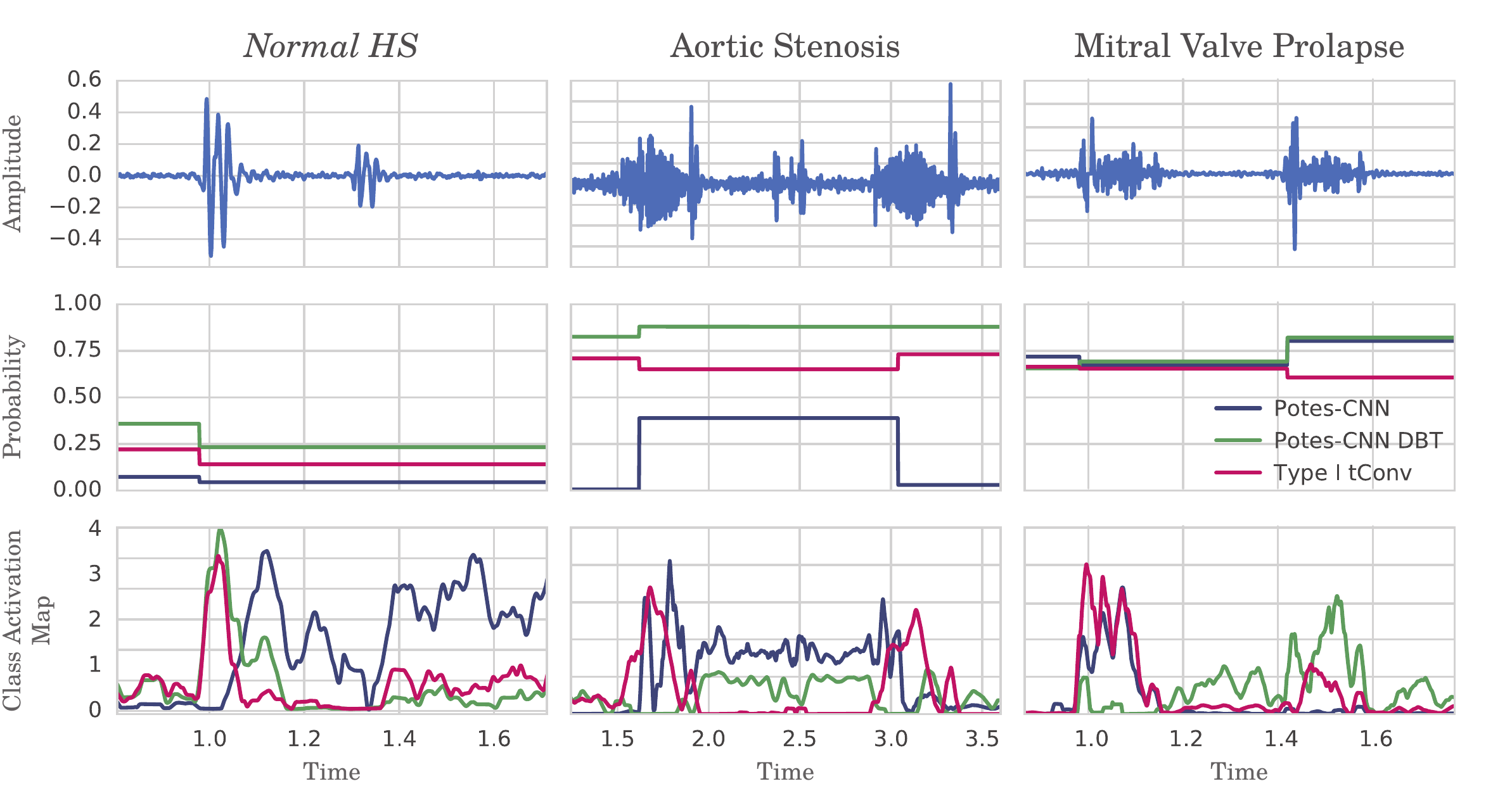}
\centering
\vspace{-3mm}
\caption{Posterior probabilities and Gradient Weighted Class Activation Maps (Grad-CAM) for a \emph{Normal} and two \emph{Abnormal} \emph{HS} segments from PHSDB. To generate Grad-CAM, activations from each branch are concatenated and weighted with the channel-wise mean of the gradients w.r.t the true class.}
\label{gradcam}
\vspace{-3mm}
\end{figure}

We can further interpret the effect of linear phase filters in the front-end by inspecting gradient weighted class activation maps (Grad-CAM) \cite{selvaraju2017grad}. Fig. \ref{gradcam} portrays the orthogonal effect of static FIR pre-processing, DBT and learnable filterbanks on the branched CNN architecture. The normal \emph{HS} segment is from subset-e while the Aortic Stenosis and Mitral Valve Prolapse \emph{HS} segments are from subset-d and subset-a respectively. The generated activation maps are smoothened with a Blackman window. For the Normal \emph{HS} segment, all of the methods give correct predictions. While Potes-CNN DBT and Type I \emph{tConv} class activation maps (CAM) are localized around the S1 and S2 \emph{HS}, the Potes-CNN method gets activated by the absence of patterns or noise in systole or diastole. For Aortic Stenosis, we can also see that the Potes-CNN method both w/ and w/o DBT, gets activated by the recording noise in the diastolic phase instead of the pathological systolic murmur. The Type-I \emph{tConv} CAM, on the other hand, is perfectly localized along with the crescendo-decrescendo aortic murmur, which is characteristic of aortic stenosis (see Fig. \ref{shapeHeart}). In the case of Mitral Valve Prolapse, the Potes-CNN method fails to localize the murmur in the second cardiac cycle, whereas Type I \emph{tConv} and Potes-CNN DBT is able to. By inspecting the CAM of our proposed learnable filterbanks and DBT scheme, we can conclude that these methods are more robust towards perturbations in the recording and indeed are being activated by the pathological markers discussed in Section \ref{murmurs}.



\subsection{Performance evaluation on HSSDB-test}
The HSSDB task is challenging by itself, considering that the best single system performance reported on this task was at $52.3\%$ (in terms of unweighted average recall) \cite{Gabor2018}.
We use this test-set to construct an unsupervised domain adaptation challenge, where the source domain is PHSDB and the target domain is HSSDB. From the results presented in Table \ref{testResults}, we observe that the proposed Gammatone \emph{tConv}-CNN and Type IV \emph{tConv}-CNN models provide the best performances in terms of Macc and F1 metrics, respectively. In particular, the gammatone \emph{tConv} front-end CNN acquires an Macc of $55.43\%$, which is superior to Gabor-BoAW-SVC by a margin of $2.41\%$ {. However, these systems were not significantly different according to McNemar's test ($p>0.01$) \cite{dietterich1998approximate}}. 
{Overall, } the results on the HSSDB-test are significantly worse compared to PHSDB-test. Although this is expected since the models were not trained on stethoscope recordings from HSSDB,  
we perform follow-up tests for further analysis.

We first test if our system performance improves when HSSDB samples are observed during training. We re-train our proposed gammatone \emph{tConv}-CNN model from scratch, using DBT while including $10\%$ of HSSDB into training randomly. This yields an Macc of $60.31\%$ on the remaining $90\%$ HSSDB recordings, with a relative improvement in Macc by $4.47\%$ compared to PHSDB-only training. Thus, the system performance indeed improves when samples from the unseen stethoscope were included during training.

To further investigate on what makes HSSDB challenging, we analyze the data using the fine-tuning method proposed in \cite{liu2019inoculation}. We fine-tune our gammatone model trained on PHSDB using 30\% data from HSSDB and evaluate its performance on the remaining 70\% HSSDB holdout data. We also monitor the change of relevant metrics in PHSDB-test. The results of the fine-tuning experiments are summarized in Table \ref{inoculation}.

In Table \ref{inoculation}, firstly, we observe that fine-tuning on 30\% HSSDB yields an increase of 1.94\% Macc on HSSDB-holdout and a decrease of 12.27\% Macc on PHSDB-test. The drop in Macc is compensated if data from both PHSDB and HSSDB is present during fine-tuning. Secondly, the specificity of our Gammatone \emph{tConv} model is reduced when HSSDB is present in training, for both PHSDB-test and the HSSDB-holdout. This indicates that distribution of \emph{Normal HS} in HSSDB is dissimilar compared to that of of PHSDB data. The sensitivity of our proposed method is increased after fine-tuning for both PHSDB and HSSDB-holdout. {Thus,} according to \cite{liu2019inoculation} (Outcome 3 in Sec. 2), we may conclude that the substandard performance is due to label noise, especially for \emph{Normal HS} in HSSDB. 
{Finally, listening to frequently misclassified \emph{Normal} samples in HSSDB-test revealed that some recordings (i.e. \texttt{devel\_0029, devel\_0056, devel\_0072}) were highly noisy with no recognizable heart sounds.}

\begin{table}[]
\centering
\caption{Performance comparison on the HSSDB-test set between baseline and proposed methods. Percentage metrics below 25\% are annotated in red.}
\label{testResults}
\resizebox{\linewidth}{!}{%
\begin{tabular}{@{}ccccc@{}}
\toprule
Method & Sens. & Spec. & Macc & F1 \\ \midrule
\multicolumn{5}{c}{\cellcolor[HTML]{EFEFEF}Baseline Systems} \\
Gabor-BoAW-SVC (Downsamp.) &  33.56 &  73.27&  53.42 & 49.28 \\
Gabor-BoAW-SVC (Upsamp.) &  28.98 & \textbf{75.86} & 53.42 & 43.27 \\
Gabor-\emph{Fusion}-SVC (Downsamp.) & 36.39 & 73.27 & 54.83 & 51.31  \\
Gabor-\emph{Fusion}-SVC (Upsamp.) & 38.69 & 62.07 & 50.38 & 52.83 \\
Potes-CNN & 63.42 & 29.31 & 46.37 & 71.3 \\
Potes-CNN-Adaboost \cite{potes2016ensemble} & 64.31 & 36.21 & 50.25 & 72.51 \\
\multicolumn{5}{c}{\cellcolor[HTML]{EFEFEF}Proposed Systems} \\
Potes-CNN DBT & 73.49 & \color{Black} 22.41 & 47.95 & 77.61 \\
Type I \emph{tConv}-CNN & 49.82 & 56.89 & 53.35 & 62.81 \\
Type II \emph{tConv}-CNN & 61.67 & 35.34 & 48.5 & 70.51 \\
Type III \emph{tConv}-CNN & 76.68 & \color{Black} 22.41 & 49.54 & 79.63 \\
Type IV \emph{tConv}-CNN & \textbf{72.08} & 26.72 & 49.41 & \textbf{77.05} \\
ZP \emph{tConv}-CNN & 73.49 & \color{Black} 23.28 & 48.38 & 77.68 \\
Gammatone \emph{tConv}-CNN & 57.42 & 53.45 & \textbf{55.43} & 68.78 \\
\bottomrule
\end{tabular}%
}
\end{table}
\begin{table}[]
\centering
\caption{Results on fine-tuning with HSSDB-test using Gammatone \emph{tConv}}
\label{inoculation}
\resizebox{\linewidth}{!}{%
\begin{tabular}{@{}ccccccc@{}}
\toprule
\multirow{2}{*}{Fine-tuning Data} & \multicolumn{3}{c|}{PHSDB-test} & \multicolumn{3}{c}{70\% HSSDB} \\ \cmidrule(l){2-7} 
 & Sens & Spec & Macc & Sens & Spec & Macc \\ \cmidrule(r){1-1}
Without retrain & 87.68 & 71.23 & 79.46 & 57.5 & 54.11 & 55.81 \\
30\% HSSDB & 89.86 & 44.52 & 67.19 & 74.55 & 41.17 & 57.87 \\
PHSDB + 30\% HSSDB & 86.95 & 63.69 & 75.32 & 61.67 & 51.76 & 56.67 \\ \bottomrule
\end{tabular}%
}
\end{table}

\section{Discussion}\label{discuss}
The presented study shows that cardiac cycle level end-to-end deep learning based methods learn features that are less affected by the domain (e.g., type of stethoscope) compared to traditional feature extraction based methods (Fig. \ref{tsne}). This is also apparent from the average subset-wise accuracy, which is considerably higher for deep learning methods compared to traditional methods. One possible explanation for this result is that short term features extracted from each cardiac cycle are less prone to domain mismatch.
As future work, this can be further examined by comparing the models with and without cardiac cycle segmentation. 
While the Gammatone \emph{tConv}-CNN is underfitting on the PHSDB-test task with the lowest subset-e accuracy (Table \ref{valResults}) and considerably higher per cardiac cycle loss compared to the other proposed methods, it was overall the best performing model among the proposed systems on the HSSDB-test set. 
This shows that auditory filterbanks could be considerably better at learning the relevant frequency bands compared to other learnable filterbanks. The Gammatone \emph{tConv} is also a non-linear phase filter with a constant group delay in its passband, enabling higher flexibility for the model coefficients. One limitation of the proposed learnable filterbanks and DBT is that it is not sufficient to learn domain agnostic features in an unsupervised manner, where the source domain and target domain are mutually exclusive. 
\section{Conclusion}
\label{conclusion}

In this study, we have proposed novel \emph{tConv} layers with a branched 1D-CNN model as learnable FIR filterbanks for stethoscope invariant heart sound abnormality detection.
Different initialization strategies and types of filter impulse responses have been examined for the \emph{tConv} layers while constraints have been added to ensure a zero and linear phase response in the resulting FIR filters. A novel gammatone filter based \emph{tConv} layer was also presented. To address the issue of unbalanced data from different stethoscopes, a training regime with a domain balanced mini-batch, termed Domain Balanced Training (DBT), was presented and its effectiveness was evaluated.
Experimental results using the proposed architecture with the DBT approach shows significant improvements compared to state-of-the-art solutions with respect to various performance metrics on multi-domain evaluation tasks prepared using the 2016 PhysioNet heart sound dataset and INTERSPEECH 2018 ComParE challenge dataset.


\ifCLASSOPTIONcaptionsoff
  \newpage
\fi

\bibliographystyle{IEEEtran}
\bibliography{refs.bib}
\end{document}